\documentclass[runningheads,a4paper,oribibl]{llncs}
\usepackage[english]{babel}
\usepackage{amssymb}
\usepackage{cite}
\usepackage{graphicx}
\usepackage{amsmath}
\usepackage{url}
\usepackage{listings}
\usepackage{amsfonts}
\usepackage{hyperref}
\newcounter{num}

\usepackage{amsmath}
\usepackage{algorithm}
\usepackage[noend]{algpseudocode}
\makeatletter
\def\BState{\State\hskip-\ALG@thistlm}
\makeatother
\numberwithin{theorem}{section}
\numberwithin{lemma}{section}
\numberwithin{definition}{section}
\numberwithin{equation}{section}
\numberwithin{figure}{section}
\numberwithin{corollary}{section}
\usepackage{color}

\definecolor{dkgreen}{rgb}{0,0.6,0}
\definecolor{gray}{rgb}{0.5,0.5,0.5}
\definecolor{mauve}{rgb}{0.58,0,0.82}

\begin{document}
\mainmatter

\title{Security of Cyber-Physical Systems}
\subtitle{From Theory to Testbeds and Validation}

 \titlerunning{Security of Cyber-Physical Systems}
\author{Jose Rubio-Hernan \and Juan Rodolfo-Mejias \and Joaquin Garcia-Alfaro}
\institute{SAMOVAR, Telecom SudParis, CNRS, Universit\'e Paris-Saclay,  Evry, France \\
\email{jose.rubio\_hernan@telecom-sudparis.eu,
  juan.mejia\_rojas@telecom-sudparis.eu, joaquin.garcia\_alfaro@telecom-sudparis.eu}
}

\maketitle
\begin{abstract}
Traditional control environments connected to physical systems are
being upgraded with novel information and communication technologies.
The resulting systems need to be adequately protected. Experimental
testbeds are crucial for the study and analysis of ongoing threats
against those resulting cyber-physical systems. The research presented
in this paper discusses some actions towards the development of a
replicable and affordable cyber-physical testbed for training and
research. The architecture of the testbed is based on real-world
components, and emulates cyber-physical scenarios commanded by SCADA
(Supervisory Control And Data Acquisition) technologies. We focus on
two representative protocols, Modbus and DNP3. The paper reports as
well the development of some adversarial scenarios, in order to
evaluate the testbed under cyber-physical threat situations. Some
detection strategies are evaluated using our proposed testbed.
\end{abstract}

\section{Introduction}
\label{sec:Introduction}

Traditional control systems are evolving in an effort to reduce
complexity and cost. These systems are converging into using a shared
network layer, enabling interconnectivity between different
manufacturers. Despite all the evident advantages of joining the
communication layer in a shared network, this evolution also opens the
door to the emergence of sophisticated cyber-threats
\cite{stuxnet,scadasec}. These threats need to be assessed to offer
novel countermeasures to minimize the risk when using shared
communication layers.

Critical services infrastructures, such as water management,
transportation of electricity, rail and air traffic control, belong to
systems nowadays coined as Cyber-Physical Systems (CPSs). The impact
of any security breach to these environments can affect the physical
integrity of individuals in contact to those systems. Even basic
threats such as replay cyber-physical attacks \cite{Teixeira2015}
could potentially cause significant damages if attack detection is not
properly undertaken. Within this scope, our goal is to put in practice
solutions of theoretical nature, modeled and implemented under
realistic scenarios, in order to analyze their effectiveness against
intentional attacks. More precisely, we assume cyber-physical
environments operated by SCADA (Supervisory Control And Data
Acquisition) technologies and industrial control protocols. We focus
on two representative protocols, which are widely used in the
industry: Modbus and DNP3 \cite{modbusspecs,dnp3specs}. Both protocols
have TCP enabled versions. This allows us the emulation of
cyber-physical environments under shared network infrastructures. We
assume a Master-Slave design pattern, which mainly dictates that
slaves would not initiate any communication unless a given master
requests an initial operation. One of our objectives has been to
combine these two protocols, both to allow the flexibility and support
of several devices with Modbus as well as the security enhancements
that DNP3 could provide as one of its features. Furthermore, some
cyber-physical detection mechanisms based on challenge-response
strategies proposed in \cite{Mo_2015,Revisiting_Wat_based_detector}
are embedded in our SCADA testbed to experiment and analyze with their
real-world performance. To complement the testbed, a set of
adversarial scenarios are designed and developed to test attacks
against the emulated environment. These scenarios focus on attacking
the Modbus segments of the SCADA architecture. The final goal is to
analyze the effectiveness of novel security methods implemented upon
the emulated environment, and under the enforcement of some attack
models.\\

\noindent {\bf Paper Organization ---} Section~\ref{sec:background}
provides the background. Section~\ref{sec:TestbedDesign} provides
details about the testbed implementation.
Section~\ref{sec:Experiments_and_Results} presents some experimental
results. Section~\ref{sec:related-work} provides related work.
Section~\ref{sec:conclusion} concludes the paper.

\section{Background}
\label{sec:background}

\subsection{SCADA Technologies}
\label{sec:CPS}

We assume Cyber-Physical Systems operated by SCADA technologies
and Industrial Control Protocols. SCADA (Supervisory Control And
Data Acquisition) technologies are composed of well-defined types of
field devices, such as: (1) Master Terminal Units (MTUs) and Human
Machine Interfaces (HMIs), located at the topmost layer and managing
device communications; (2) Remote Terminal Units (RTUs) and
Programmable Logic Controllers (PLCs), controlling and acquiring data
from remote equipment and connecting with the master stations; and (3)
sensors and actuators.

The MTUs of a SCADA system are located at the control center of the
organization. The MTUs give access to the management of
communications, data storage, and control of sensors and actuators
connected to RTUs. The interface to the administrators is provided via
the HMIs. The RTUs are stand-alone data acquisition and control units.
Their tasks are twofold: (1)~to control and acquire data from process
equipment (at the remote sites); and (2)~to communicate the collected
data to a master (supervision) station. Modern RTUs may also
communicate between them (either via wired or wireless networks). The
PLCs are small industrial microprocessor-based computers. The
significant differences with respect to an RTU are in size and
capability. Sensors are monitoring devices responsible for retrieving
measurements related to specific physical phenomena, and communicate
such measurements to the controllers. Actuators translate control
signals to actions that are needed to correct the dynamics of the
system, via the RTUs and PLCs.

\subsection{Industrial Control Protocols}
\label{sec:protocols}

Protocols for industrial control systems built upon SCADA technologies
must cover regulation rules such as delays and faults \cite{brown2000overview}.
However, few protocols imposed by industrial standards provide
security features in the traditional ICT security sense. Details
about two representative SCADA protocols used in our work follow.

{\bf Modbus --} One of the first protocols that stands out when working
with data acquisition systems is Modbus \cite{modbusspecs}. It was
developed around the 80's and it was done with no security concerns as
was common at that time. It was developed by Modicon to be used with
their PLCs. The protocol was formulated as a method to transmit data
between electrical devices over serial lines. In the standard working
mode, Modbus has a master and slave architecture, something really
common for half duplex communications. The protocol is free and open
source, making it really popular among the automation industry. The
protocol evolved to allow different communication technologies. For
instance, Modbus ASCII, for serial communications; and Modbus TCP/IP
for Ethernet networks.

{\bf Distributed Network Protocol (DNP3) --} As with Modbus, DNP3 is a
query-response protocol for process automation systems. Messages are
sent over serial bus connections or Ethernet networks (using the
TCP/IP stack) \cite{dnp3specs}. The protocol recently has been leaning
towards a more security-oriented design. Previous versions of the
protocol suffered from the same kind of design conception where
security was not taken into account, due to the inherent level of
security that dedicated networks provided by this protocol.

\subsection{Control-theoretic Protection}
\label{sec:Simulation_Testbed}

Cyber-physical systems operated by SCADA technologies and industrial
control protocols can be represented as closed-loop systems. Such
systems follow closely the pattern of controlling the system based on
the feedback they are getting from measurements. Several
control-theoretic solutions have been presented in the literature to
detect attacks against cyber-physical systems. In \cite{wu2016survey},
some techniques are presented to improve the security of networked
control systems using control theory. A proper example is the use of
\emph{authentication watermarks}. Stationary watermarks, i.e.,
Gaussian zero-mean distributed signals, are added to the control
signals, in order to identify integrity attacks against the system.
The watermark-based detector can identify the effect of real
distribution values, generated at the output, with regard to, e.g.,
replayed or injected distribution values \cite{Mo_2015}. However,
adversaries with enough resources to infer the dynamics of the
protected system can evade detection
\cite{Revisiting_Wat_based_detector}. Indeed, there are several
methods that a potential attacker can use to identify and learn the
behavior of the system \cite{sysid1}. The goal of these techniques is
to obtain a mathematical model of the system, based on eavesdropped
measurements. Non-parametric system identification techniques include
the use of adaptive filters, such as Finite Impulse Response (FIR)
filters. Some more powerful techniques to identify complex dynamic
systems include the use of autoregressive methods, such as ARX
(autoregressive exogenous model) and ARMAX
(autoregressive-moving-average model with exogenous inputs model)
\cite{sysid2}. These techniques can be used by malicious adversaries,
in order to estimate the parameters of the system prior executing
their attacks. Improvements to address those aforementioned problems
have been presented in
\cite{Revisiting_Wat_based_detector,controlsys}. The goal of the
testbed presented in the following section is to validate the
effectiveness of the aforementioned techniques. Data derived from the
testbed is expected to complement theoretical and numeric simulations
provided in previous work.

\section{Testbed Design}
\label{sec:TestbedDesign}

\begin{figure}[!b]
\centering
\includegraphics[width=0.7\columnwidth]{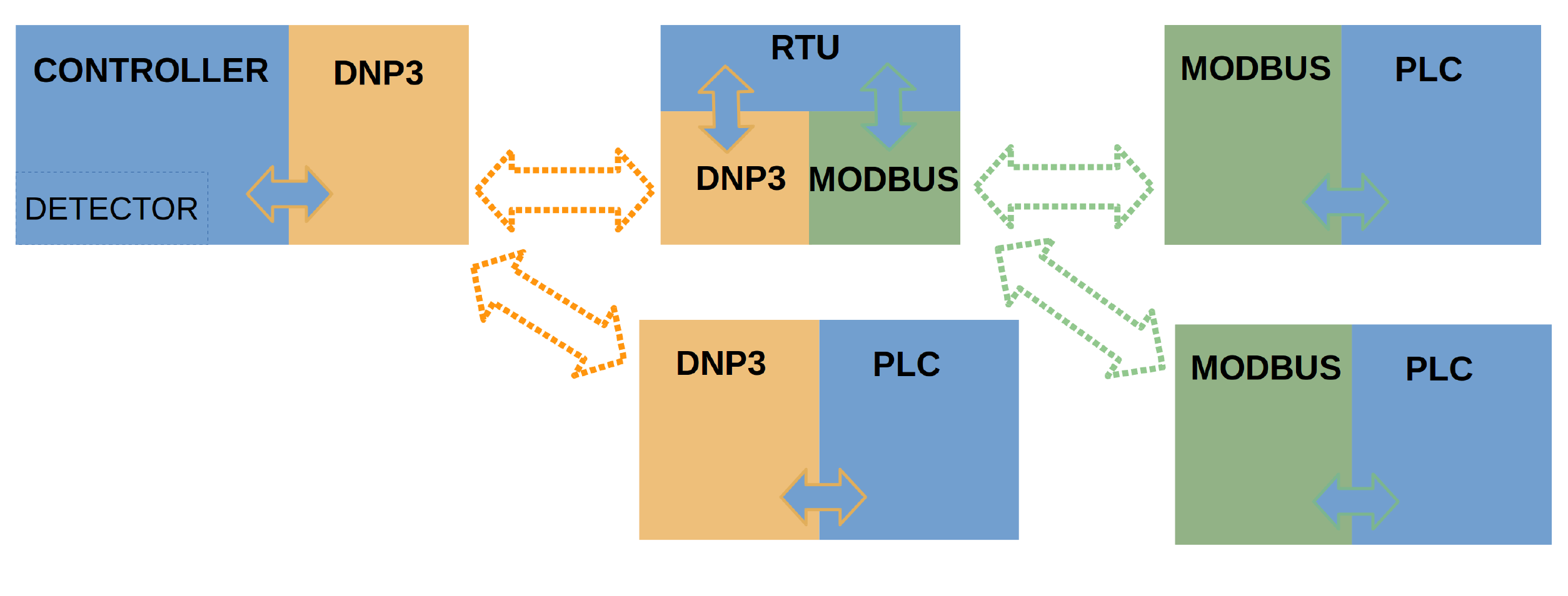}
\caption[Architecture]{Abstract architecture overview.}
\label{fig:archdesign}
\end{figure}

\subsection{Architecture}
\label{sec:archdesign}
Closed-loop systems are systems which rely upon internally gathered
information to perform, correct, change or even stop actions. This
kind of systems are important in the control theory branch, known to
have two-way communication, one to read data and the other to forward
commands.

We can observe three important block elements: the controller, the
system itself, and the sensors. The controller reads data from the
sensors, computes new information and transmit new commands to the
system (i.e., the system control input). The system control input is
generated by the controller with the purpose of correcting the
behavior of the system, under some previously established limits. The
system is what we normally see as the entity under control. The
sensors are the feedback link between the system and the controller.
Their purpose is to quantify the output and provide the necessary
information to the controller, in order to compare and, if necessary,
correct the behavior of the system.

The architecture proposed for our SCADA testbed works as follows. All
the aforementioned elements can be distributed across several nodes in
a shared network combining DNP3 and Modbus protocols (cf. Figure
\ref{fig:archdesign}). Likewise, one or various elements can be
embedded into a single device. From a software standpoint, the
controller never connects directly to the sensors. Instead, it is
integrated in the architecture as a SCADA PLC (Programmable Logic
Controller) node, with eventual connections to some other
intermediary nodes. Such nodes are able to translate the controller
commands into SCADA (e.g., either Modbus or DNP3) commands. As
depicted in Figure \ref{fig:archdesign}, the architecture is able to
handle several industrial protocols and connect to complementary SCADA
elements, such as additional PLCs and
RTUs (Remote Terminal Units). To evolve the architecture into a
complete testbed, new elements can be included in the system, such as
additional proxy-like RTU nodes.

From a data transmission standpoint, we include in our SCADA testbed
the possibility of using different sampling frequencies, in order to
cover a larger number of experimental scenarios. The implementation is
based on control theory \cite{controlsys} supporting the use of
different frequencies when performing read and write operations.
Specifically, this narrows to the sampling frequency, a system with
mono-frequency sampling is where the same frequency is used for all
the channels or multi-frequency sampling where different sampling
frequencies are used in each channel. Depending on the nature of the
system mono or multi is better.

The architecture is able to handle many PLCs. To avoid overloading one
channel with all the possible registers of the PLCs, separate ports
are designated in order to isolate the communication between separated
PLCs. DNP3 commands perform an Integrity Scan which gathers all the
data from the PLCs in case several PLCs were being handled in the same
channel, all variables of the a PLC would be fetched causing overhead
in the communication.

\subsection{Implementation Design}
\label{sec:implementation}

\begin{figure}[!b]
\centering
\includegraphics[width=0.8\columnwidth]{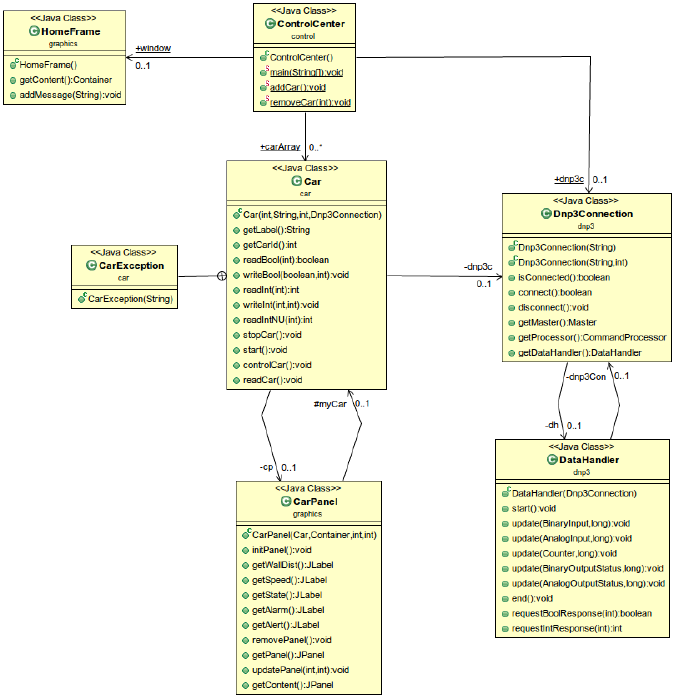}
\caption[Controller Class Diagram]{Implementation overview, controller side.}
\label{fig:class_diagram}
\end{figure}

The implementation of our SCADA testbed consists on {\it Lego
  Mindstorms} EV3 bricks \cite{legoev3} and Raspberry Pi \cite{rpi}
boards as PLCs to control some representative sensors (e.g., distance
sensors) and actuators (e.g., speed actuators). We refer the reader to
\url{http://j.mp/legoscada} for additional information. Figure
\ref{fig:class_diagram} shows an object-oriented representation of the
testbed implementation, along with connection control classes,
exception classes and also graphical interface classes at the
controller side. In Figure \ref{fig:rtu_diagram}, we can see all the
classes that have been created in order to achieve the DNP3-Modbus
combination, at the RTU side. A proxy-like behavior has been also
implemented allowing to translate the commands in both directions for
both protocols.
\subsubsection{Controller Design ---}
\label{sec:controllerDesign}

The controller has a graphical interface to show the behavior of the
system to an operator. It is orchestrated by the {\it ControlCenter}
class (cf. Figure \ref{fig:class_diagram}). This class handles the
graphical interface (cf. {\it HomeFrame} class) representing the HMI
(Human Machine Interface) of the SCADA architecture. Some PLC
instances, e.g., the {\it Car instances}, subsequently create DNP3
connections under a {\it DataHandler}, which is in charge of managing
the communications between RTUs and PLCs. Finally, some of the
instances (e.g., the {\it Car instance}) implement a graphical
component to provide additional information to the operator.

\subsubsection{RTU Design ---}
\label{sec:RTUDesign}

\begin{figure}[!t]
\centering
\includegraphics[width=0.7\columnwidth]{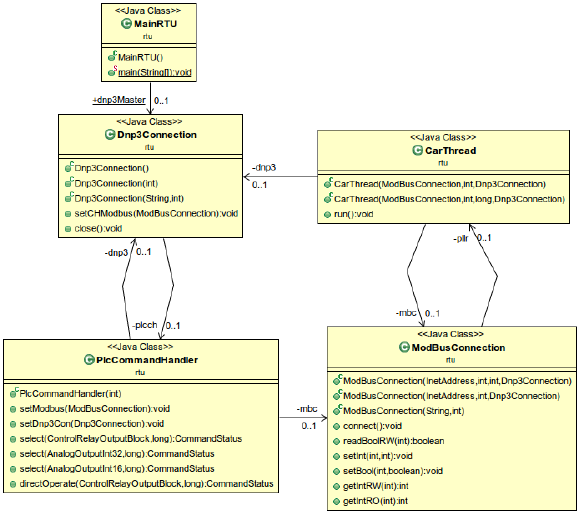}
\caption[RTU Class Diagram]{Implementation overview, RTU side.}
\label{fig:rtu_diagram}
\end{figure}

In the implementation, it is possible to have control of one or more
PLC instances. For such a task, a dedicated thread manages the
translations and constant polling of each PLC. Everything starts with
the { \it MainRTU} class (cf. Figure \ref{fig:rtu_diagram}), which
opens the main DNP3 connection to expect the controller. Once the
controller connects, the RTUs exchange information of the PLCs to add,
and create all the respective classes in order to handle each PLC
individually and with dedicated ports.

\begin{figure}[!b]
\centering
\includegraphics[width=0.7\columnwidth]{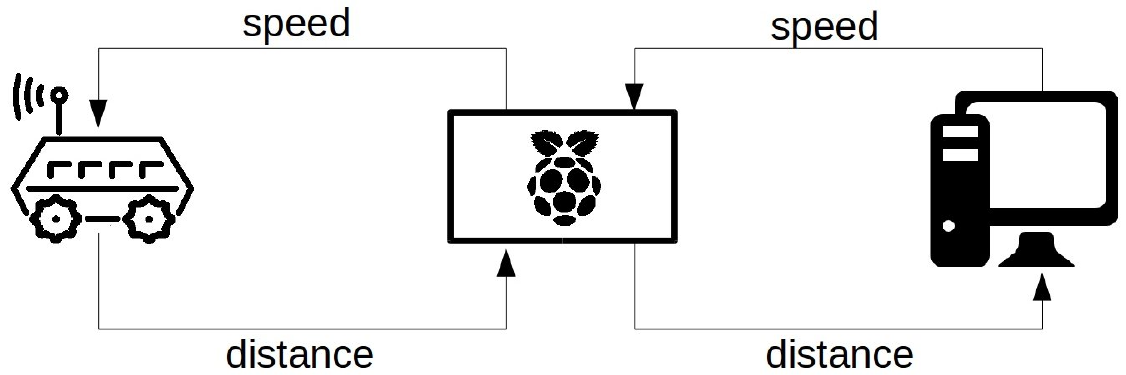}
\caption[Test Scenario]{Test scenario overview (cf. \url{http://j.mp/legoscada}).}
\label{fig:archTestdescription}
\end{figure}

\subsubsection{External Tools ---}
\label{sec:External_Implemented}

Apart from the architecture implementation, other tools have been
implemented in order to facilitate the aggregation process of new
SCADA nodes. Specific custom scripts have been made to install the
OpenDNP3 libraries either compiling them from source or use
precompiled binaries for the case of the Raspberry Pi. Compiling
source is time-consuming if it is done directly at the Raspberry Pi
boards. Therefore, cross-compiling or precompiled libraries are
recommended to avoid long compilation times. The Raspbian scripts give
the choice to use precompiled libraries.

\subsection{Test Scenario Description}
\label{sec:Test_Scenario}

Figure \ref{fig:archTestdescription} shows the components of a test
scenario. This scenario is a simple representation of the architecture
proposed in this paper. It consists of a controller (Personal
Computer), an RTU (Raspberry Pi) and a PLC (Lego EV3 Brick). The
controller is always correcting the car speed and polling the distance
between the car and an obstacle. One single controller and one single
RTU can control various PLCs. To start the testbed is necessary to
launch the Java program on the brick \cite{jamod}, and the
intermediary Java software in the Raspberry Pi board. When starting
the controller and adding a car, the controller communicates with each
layer to perform the request. The car behavior is continually being
modified by the controller hence varying the car speed many times per
second.

\subsection{Implementation of MiM Attacks}
\label{sec:attackerimple}

Man-in-the-middle (MiM) is a very common type of attack, compromising
the communication in both ends, especially if the communication is not
encrypted as is the case with many SCADA implementations. As mentioned
previously, the watermark detectors reported in
\cite{Revisiting_Wat_based_detector,controlsys} are implemented in our
testbed. Theoretical proofs and numeric simulations were already
conducted to validate the proposed detectors. The testbed proposed in
this paper was expected to provide complementary validation of the
detectors. After having an entire architecture working, the next
requirement was to implement the adversarial scenarios reported in
\cite{Revisiting_Wat_based_detector,controlsys}.

In order to develop the scenarios, an attacker model was used as a
base for assumptions to define the opponents' capabilities. We assume
that the attacker can intercept all communication between ends, and
thus the attacker can alter, store, analyze replay and forge false
data in the communication. Since this is done using a testbed instead
of numeric simulations, all real-life limitations are applied to the
attacker. ARP poisoning \cite{nam2010enhanced} is used by the attacker
to intercept the channels and eavesdrop the communications. The
attacker has a passive and active mode of operation. The {\it passive
  mode} is where the attacker only eavesdrops, processes, and analyzes
the data without modifying the information contained in the payload of
the messages. Nevertheless, Ethernet header data, such as the hardware
addresses, are modified since ARP tables are poisoned. During the {\it
  active mode}, the attacker starts injecting data to the hijacked
communication. This injection, depending on the pattern of the
attacker, can be a generated response or replayed packets.

\subsubsection{Replay Attack ---}
\label{sec:replay}
The attacker uses ARP poisoning to start eavesdropping the connection
(passive mode). After capturing enough data, the {\it active mode}
starts. The attacker injects the old captured data following the
stream of packets of the previous capture. Before starting to disrupt
the system, the attacker conducts the attack between the sensors and
the controller, forging only the TCP headers that correspond to the
opened TCP sessions. Once replayed the packets, the system gets
disrupted by forging data between the controller and the PLCs.

\subsubsection{Injection Attacks ---}
\label{sec:model_inferring_attackers}
Prior to starting the attacks, the attacker eavesdrops connections
using the {\it passive mode}, and analyze the data in order to infer
the dynamics of the system. This is used to evade the authentication
watermark detector. Once inferred the model of the system, the
attacker starts injecting correct data in the communication in order
to defeat the watermark countermeasure. To delude the detector, the
attacker calculates the effect of the watermark in the system and
tries to cancel the ability of the detector to sense the changes in
the feedback signal. Two different techniques are implemented: 1) a
non-parametric filter, called Finite Impulse Response filter (FIR), in
order to implement the evasion technique presented in
\cite{Revisiting_Wat_based_detector}; and 2) autoregressive methods,
such as ARX and ARMAX, in order to implement the evasion technique
presented in \cite{controlsys}.

\subsection{Attack and Fault Detection}
\label{sec:detector}

The adaptation of a fault detector in order to detect attacks using an
authentication watermark is a valid technique that has been proved to
work in \cite{Mo_2015,Revisiting_Wat_based_detector,controlsys}. The
aforementioned techniques have been implemented in our SCADA testbed,
in order to assess and analyze their performance using real hardware
components. The testbed controllers have built-in the detector with
different types of watermarks
(cf.~Section~\ref{sec:Simulation_Testbed}). The implementation uses
the {\it JKalman} library \cite{jkalman}, with some light
modifications to parameterize the system and detect the effect of the
watermark in the system's output. The detector estimates the next
output of the system and then compares it to the value returned by the
system. The process uses the $\chi^2$ detector proposed in
\cite{Mo_2015}. The detector returns a metric, $g_{t}$, which
increases rapidly when the output of the system starts to move away
from the estimation. The metric is posteriorly used to generate
alerts.

The $g_{t}$ metric is an in-code operator that quantifies the
difference between the parametric model output and the actual system
output. An increase of $g_{t}$ means that the system is not behaving
or reacting to the watermark as expected. Therefore, the system is
likely to be under attack. The value of $g_{t}$ is calculated for each
iteration and compared with the values of some previous iterations. In
order to discard false positives, the controller implements the
validation code presented in Algorithm \ref{f_A_diff}, to separate
normal faults from attacks or severe failures. The algorithm alerts
the operator only when real intervention is required, making the
differentiation between faults, e.g., latency or inaccuracy events at
the sensor; and intentional attacks. For every feedback sample, the
controller analyzes $g_{t}$. If $g_{t}$ consecutively bypasses a given
threshold more than $window$ times, then it triggers an $alert$.
\begin{algorithm}[ht]
\caption{--- Fault and Attack Detector}
\label{f_A_diff}
\begin{algorithmic}[1]
\Procedure{detection algorithm}{}
\State $\textit{window} \gets \textit{detector window}$
\BState \emph{loop}:
\If {$g_{t} \geq threshold$}
	\State $risk\gets risk+1$.
    \If {$risk > window$}
    	\State $alert\gets alert+1$.
    \EndIf
\Else
	\State $risk\gets 0$.
\EndIf
\EndProcedure
\end{algorithmic}
\end{algorithm}

\section{Experimentation and Results}
\label{sec:Experiments_and_Results}

\subsection{Experimentation}
\label{sec:experiments}

We present in this section the results of applying the watermark
authentication technique presented in
\cite{Revisiting_Wat_based_detector,controlsys} under the testbed
presented in Section~\ref{sec:TestbedDesign}. Several repetitions of
the experiment were orchestrated using automated scripts handling the
elements of our SCADA testbed scenarios. The scripts can perform
several actions, such as starting the controller and the RTUs, as well
as executing the predefined attacks. A set of attacks and detectors
have been used and posteriorly analyzed. The combinations,
attack--detector, are the following:

\begin{itemize}
\item \emph{Replay Attack--Watermark Disabled:} the attacker is
  likely to evade the detector, since no watermark is injected into
  the system.

\item \emph{Replay Attack--Watermark Enabled:} the attacker is
  likely identified by the detector, since the attack is not able to
  adapt to the current watermark.

\item \emph{Non-parametric Attack--Stationary Watermark:} attacker
  and detector have equal chances of success.

\item \emph{Non-parametric Attack--Non-stationary Watermark:} the
  non-stationary watermark changes the distribution systematically,
  hence preventing the FIR-based attack to adapt to such changes. The
  expected results are an increase of the detection ratio.

\item \emph{Parametric Attack--Stationary Watermark:} the attacker
  is likely to evade the detector when the attack properly infers the
  system parameters.

\item \emph{Parametric Attack--Non-stationary Watermark:} the
  attacker is also likely to evade the detector when the system
  parameters are properly identified.
\end{itemize}

The cyber-physical implications of the testbed hinder the
experimentation process especially when several repetitions are
required in order to obtain statistical results, contrary to
simulations where only the code is executed. The creation of the
orchestration script, which automates the test, has been necessary
to simplify the experimentation tasks. Next section shows the results
using the testbed for the aforementioned
attacker-detector combinations. A sample execution of the {\it Replay Attack --
  Watermark Disabled} scenario is available at
\url{http://j.mp/legoscada}.

\subsection{Experimental Results}
\label{sec:results}

\begin{figure}[H]
  \begin{minipage}[t]{.48\textwidth}
    \includegraphics[width=.98\columnwidth]{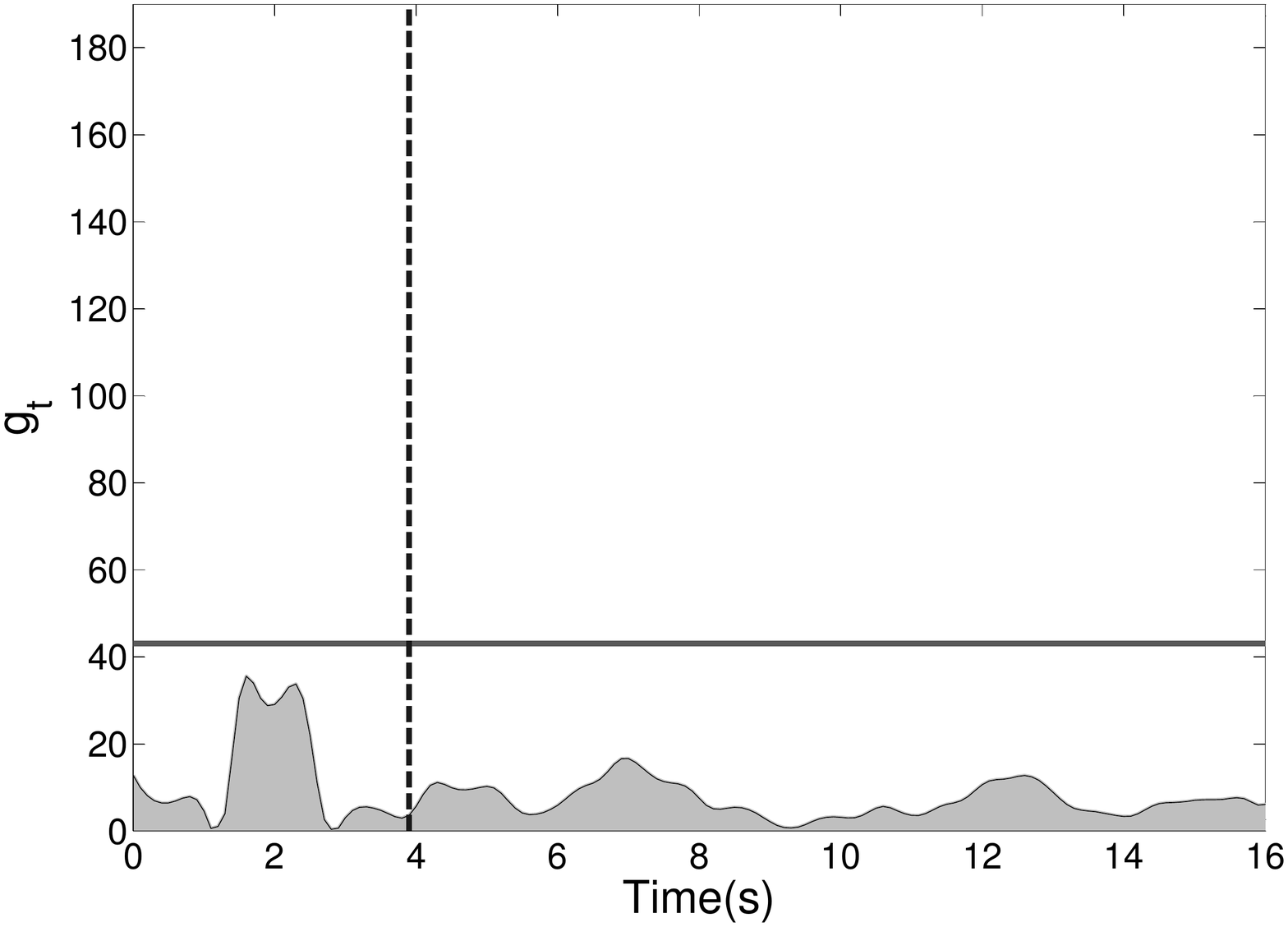}
    \\(a) No watermark under replay attack\\~~\\
  \end{minipage}
  \hfill
  \begin{minipage}[t]{.48\textwidth}
    \includegraphics[width=.98\columnwidth]{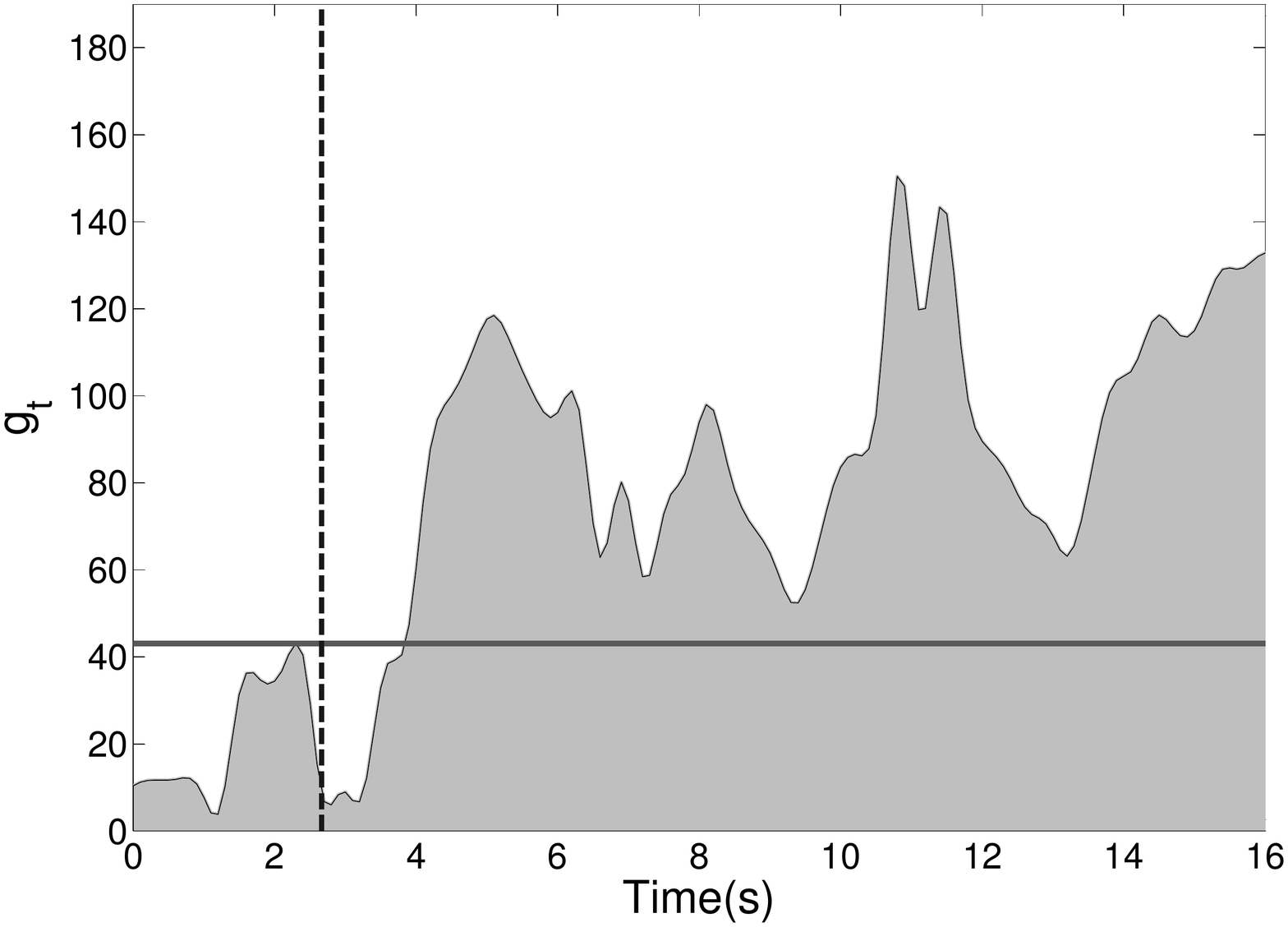}
    \\(b) Stationary watermark under replay attack \\~~\\
  \end{minipage}
  \begin{minipage}[t]{0.48\textwidth}
    \includegraphics[width=.98\columnwidth]{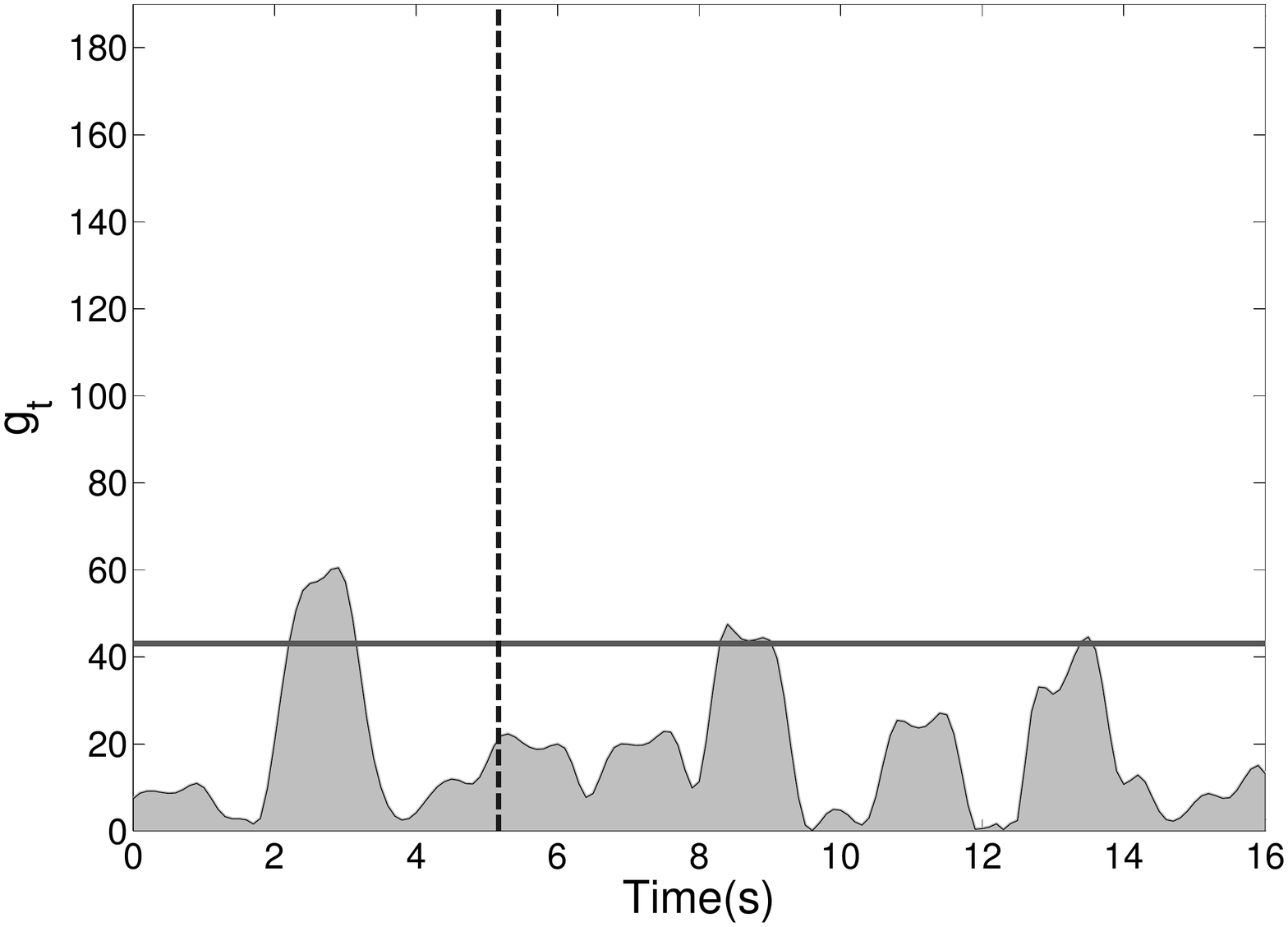}
    \\(c) Stationary watermark under non-parametric attack\\~~\\
  \end{minipage}
  \hfill
  \begin{minipage}[t]{0.48\textwidth}
    \includegraphics[width=.98\columnwidth]{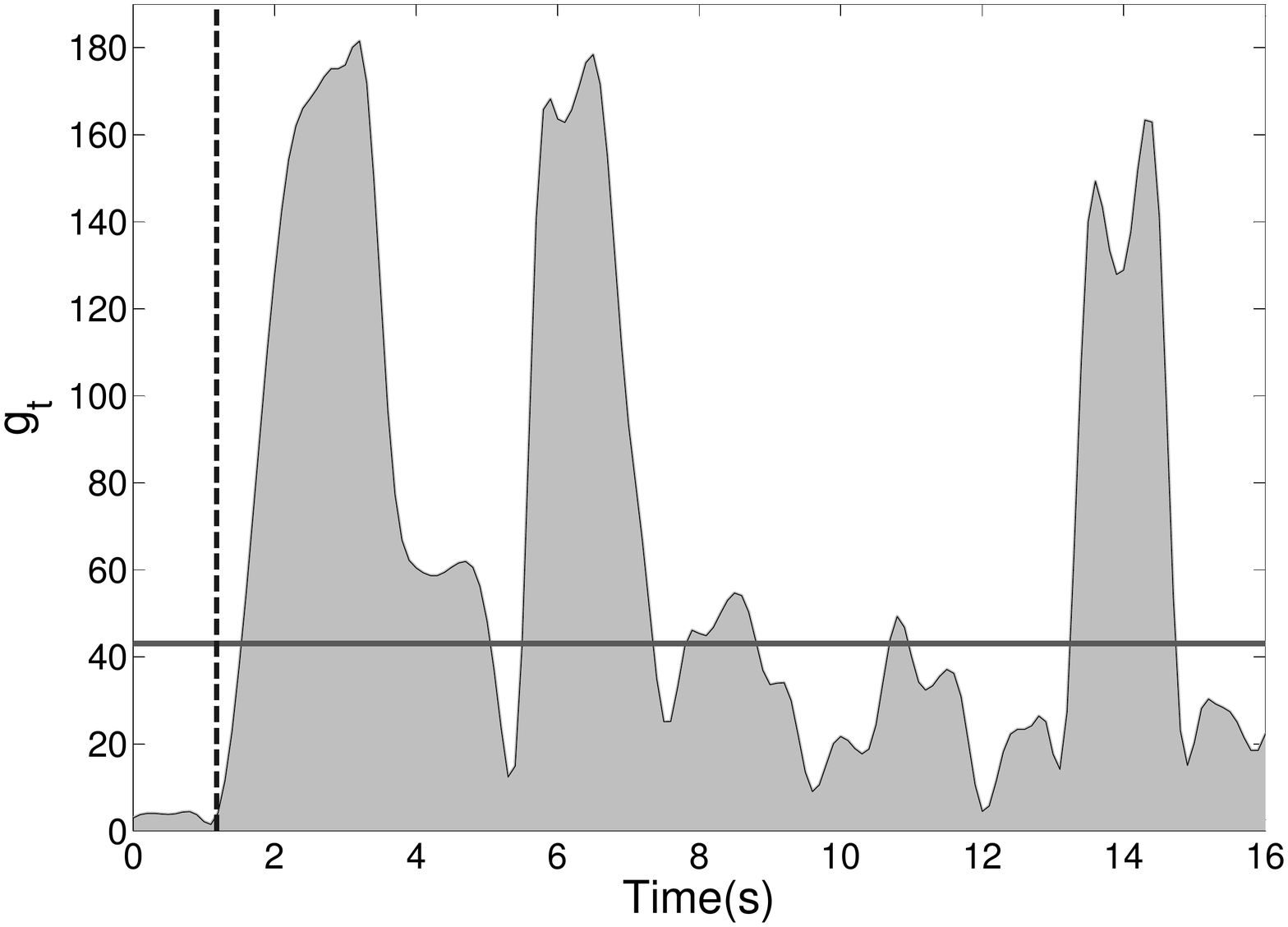}
    \\(d) Non-stationary watermark under non-parametric attack\\~~\\
  \end{minipage}
  \begin{minipage}[t]{.48\textwidth}
    \includegraphics[width=1\columnwidth]{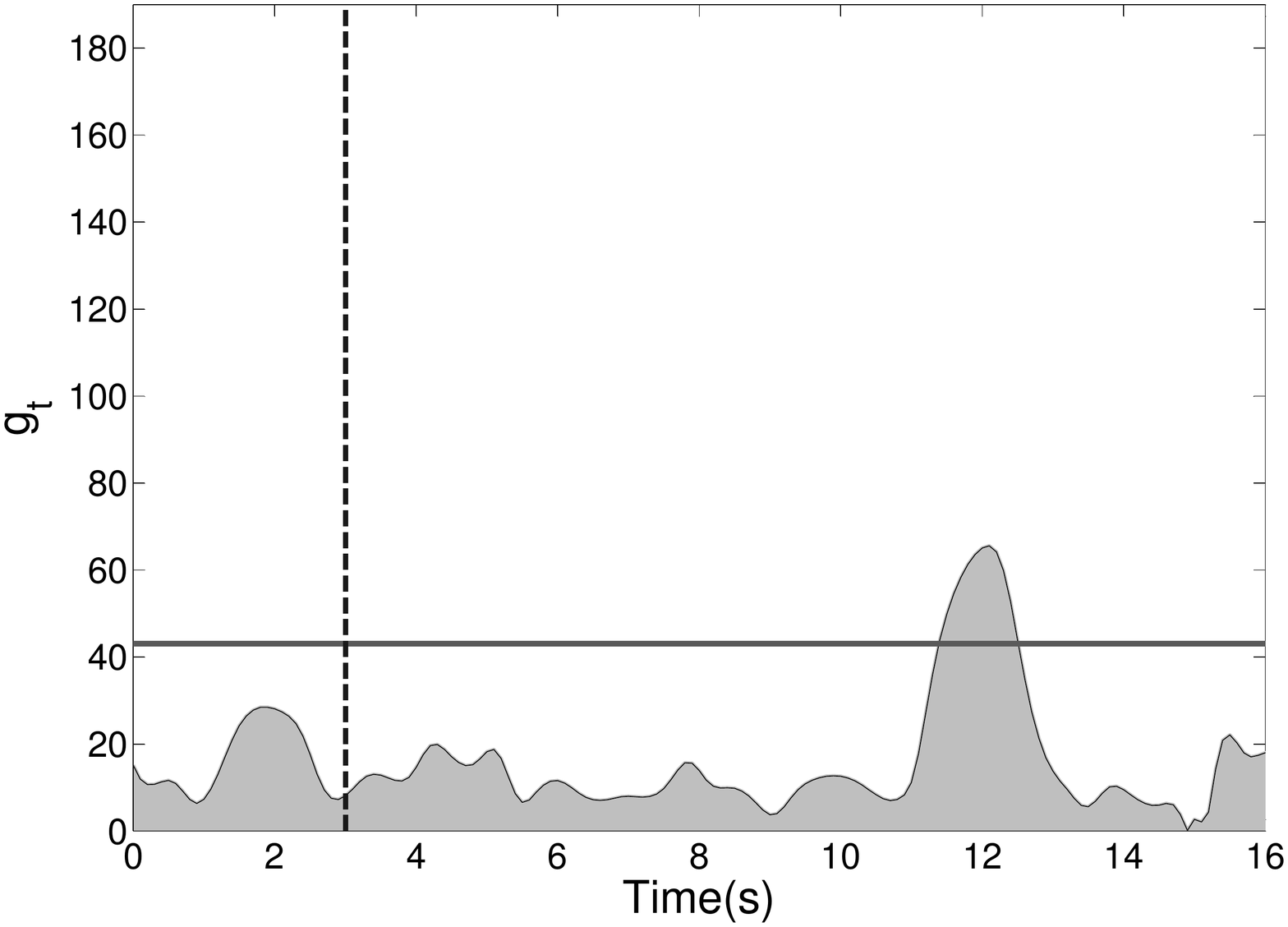}
    \\(e) Stationary watermark under parametric attack\\~~\\
  \end{minipage}
  \hfill
  \begin{minipage}[t]{.48\textwidth}
    \includegraphics[width=1\columnwidth]{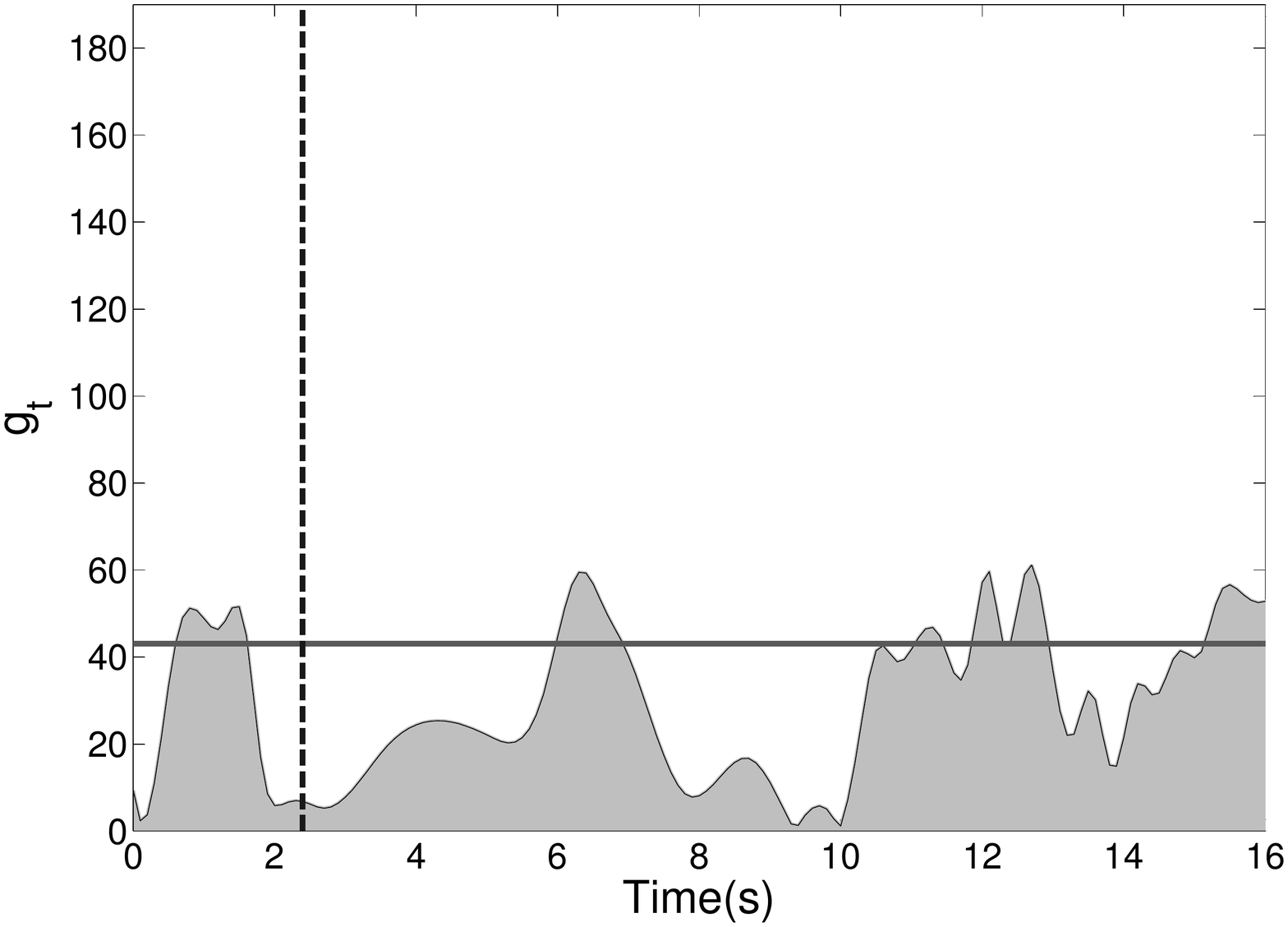}
    \\(f) Non-stationary watermark under parametric attack\\~~\\
  \end{minipage}
  \caption[Results]{Detection results. The horizontal solid line
    represents the threshold. The vertical dotted line represents the
    moment when the attack starts. Peaks on the left side of the
    vertical dotted line represent false positives. (a),(b) detection
    values of $g_{t}$, without and with stationary watermark under
    replay attack. (c),(d) detection values with stationary and
    non-stationary watermark under non-parametric attack. And (e),(f)
    detection values with stationary and non-stationary watermark
    under parametric attack.}
\label{fig:detOutput}
\end{figure}

After collecting data from different devices across the SCADA testbed,
the data is analyzed accordingly to interpret the performance of the
detector with regard to the attack scenario. Since the stationary watermark
detector was correctly refined for each test scenario, we are able to
analyze in depth the results through a statistical evaluation of the
data. Experimental results with the non-stationary watermark mechanism
are also conducted. Figure~\ref{fig:detOutput} shows the detector
values, $g_{t}$, for all the attack-detector combinations defined in
Section~\ref{sec:experiments}.

For all the plots, the solid horizontal line represents the threshold;
and the vertical dotted line represents the moment when the attacker
starts injecting malicious data. The short peaks on the left side of
the plots, those bypassing the threshold line before the start of the
attacks, are counted as false positives or system faults.

Figures~\ref{fig:detOutput}(a) and \ref{fig:detOutput}(b) show the
experimental results of the replay attack. When the watermark was
disabled (cf. Figure~\ref{fig:detOutput}(a)), the attacker properly
evades the detector. Since the controller is not inserting the
protection watermark, it does not detect the attack. On the contrary,
the results in Figure~\ref{fig:detOutput}(b) show that the activation
of the watermark under the same scenario allows the controller to
alert about the attack almost immediately. Based on these results, we
can conclude that the stationary watermark based detector properly works out to detect the replay attack.

Figure \ref{fig:detOutput}(c) represents the non-parametric attacker
against the previously tested stationary watermark. The detector is
now unable to detect the attacker. Figure~\ref{fig:detOutput}(d) shows
the case where the non-stationary watermark is enabled. Under this
situation, the detector has lightly more chances of detecting the
attack. This shows how the non-stationary watermark mechanism does
improve the detection abilities compared to the stationary watermark
approach.

Figures~\ref{fig:detOutput}(e) and~\ref{fig:detOutput}(f) evaluate the
scenario associated to the parametric attacks. Theoretically, the
attacker is expected to evade the detector when the attack succeeds at
properly identifying the parameters of the system dynamics.
Figure~\ref{fig:detOutput}(e) represents the experiments where the
parametric attack is executed under the stationary watermark scenario.
The figure shows that the detector value, $g_{t}$, remains most of the
time below the detection threshold. Figure~\ref{fig:detOutput}(f)
shows the behavior of the detector under the non-stationary watermark
scenario. This time, the detector has slightly more chances of
detecting the attack.

\subsection{Statistical Data Evaluation}

Using the watermark-based detection mechanism, we run for each attack
scenario 75 automated rounds (about 4 hours of data collection
processing). In order to evaluate the results, we use the following metrics:

\begin{enumerate}
\item {\it Detection Ratio}, associated to the success
  percentage of the detector, calculated with regard to time range
  after each attack starts.

\item {\it Average Detection Time}, determining the amount of time
  needed by the detector to trigger the attack alert.

\item {\it False Negative (FN)} ratio, determining the number of
  samples where the detector fails at successfully alerting about the
  attacks. The ratio is calculated as follows,
\begin{equation}
\label{eq:FN}
  FN=\frac{SA-AD}{SA}
\end{equation}
where \emph{SA} represents the values of the samples under attack, and
\emph{AD} the samples detected as an attack.

\item {\it False Positives (FP)} ratio, calculated as the number
  of samples where the detector signals benign events as attacks. The
  ratio is calculated as follows,
\begin{equation}
\label{eq:FP}
   FP=\frac{AD}{SN}
\end{equation}
where \emph{SN} represents the number of samples under normal
operation, and \emph{AD} the number of samples detected as attack by
mistake.
\end{enumerate}

Table \ref{tab:statisicratio} shows the performance results of the
detector, based on the \emph{Detection Ratio} and the \emph{Average detection Time} metrics.
\vspace{-0.4cm}
\begin{table}[H]
\centering
  \begin{tabular}{|c|c|c|c|}
    \hline
        & {\bf  Replay Attack}  & \begin{tabular} {@{}c@{}} {\bf  Non-parametric }  \\ {\bf Attack}  \end{tabular}  & \begin{tabular} {@{}c@{}} {\bf  Parametric }  \\ {\bf Attack}  \end{tabular}\\
    \hline
    {\it Detection Ratio}  &   40.00\%  &  18.00\% &  12.00\%   \\
    \hline
     {\it Average Detection Time}  &   18.81s &  10.17s &  6.08s  \\
    \hline
  \end{tabular}
\medskip
\caption[Detector Performance]{Detector performance results.}
\label{tab:statisicratio}
\end{table}
\vspace{-0.5cm}

Regarding the results shown in Table \ref{tab:statisicratio}, we can emphasize that the replay attack is the most detectable scenario, with a detection ratio of about $40\%$. This detection ratio is still far from perfect, maybe due to
the sensors accuracy and resolution; but better than for the rest of
scenarios. The non-parametric attacker has a lower detection
ratio, of about $18\%$. This result is expected, as suggested by the
theoretical and simulation-based conclusions available at
\cite{Revisiting_Wat_based_detector}, where the authors emphasize that
the mechanism is not sufficiently robust to detect adversaries that
are able to identify the system model. To finish, the parametric
attack has the most robust system identification approach. The
attacks can evade the detection process if they succeed at properly
identifying the system attributes. In terms of results, they lead to
the lowest detection rate of about $12\%$.

During the replay attack, the {\it Average Detection Time} is the
slowest of all the adversarial scenarios. This behavior is due to the
watermark distribution properties, since the watermark variation makes
the replay attack highly detectable. At the same time, the injection
attacks (either the parametric or the non-parametric version) are
detected much faster than the replay attack. This is due to the
transition period needed by the attackers to estimate the correct data
prior misleading the detector. For this reason, if the attacker does
not choose the precise moment to start the attack, the detector
implemented at the controller side is able to detect the injected
data, right at the beginning of the attack. Furthermore, the attackers
shall also synchronize their estimations to the measurements sent by
the sensors. In case of failing the synchronization process, the
detector does identify the uncorrelated data, and reports the attack.

Table \ref{tab:statisicresults} shows that the detection of the replay
attack has the lowest false negative ratio, $64.06\%$, hence
confirming that this adversarial scenario is the most detectable
situation with regard to the detection techniques reported in
\cite{Mo_2015}. The detection of the non-parametric attacks has a
higher false negative ratio, $85.20\%$, confirming the theoretical and
simulation-based results reported in
\cite{Revisiting_Wat_based_detector}. The detection of the parametric
attacks also confirms the results estimated in \cite{controlsys}, and
leading to the highest false negative ratio, $88.63\%$. Finally, and
in terms of false positive ratio, the three adversarial scenarios show
a low impact (on average, about $1.33\%$ false positive ratio). Such
low impact is, moreover, easy to tune by adapting the parameters of Algorithm
\ref{f_A_diff}.

\vspace{-0.4cm}
\begin{table}
\centering
  \begin{tabular}{|c|c|c|c|}
    \hline
        &  {\bf  Replay Attack}   & \begin{tabular} {@{}c@{}} {\bf  Non-parametric }  \\ {\bf Attack}  \end{tabular}  & \begin{tabular} {@{}c@{}} {\bf  Parametric }  \\ {\bf Attack}  \end{tabular}\\
    \hline
     {\it False Negatives }  &   64.06\% &  85.20\% &  88.63\%  \\
     \hline
     {\it False Positives }  &    0.98\%    &  1.66\% &  1.35\%   \\
    \hline
  \end{tabular}
\medskip
\caption[Statistical Results]{Long run experiment results.}
\label{tab:statisicresults}
\end{table}
\vspace{-1.1cm}

\section{Related Work}
\label{sec:related-work}

The study of security incidents associated to cyber-physical systems
underlying critical infrastructures has gathered a big amount of
attention since the infamous Stuxnet case \cite{stuxnet}. Since then,
research on cyber-physical systems has progressed substantially
resulting in a large number of testbeds developed and established
in the literature. A non-exhaustive list follows.

Myat-Aung present in \cite{itrust} a Secure Water Treatment (SWaT)
simulation and testbed to test defense mechanisms against a variety of
attacks. Siaterlis et al. \cite{siaterlis2013epic} define a
cyber-physical Experimentation Platform for Internet Contingencies
(EPIC) that is able to study multiple independent infrastructures and
to provide information about the propagation of faults and
disruptions. Green et al. \cite{green2016testbed} focus their work on
an adaptive cyber-physical testbed where they include different
equipments, diverse networks, and also business processes. Yardley
reports in \cite{yardley_testbed} a cyber-physical testbed based on
commercial tools in order to experimentally validate emerging research
and technologies. The testbed combines emulation, simulation, and real
hardware to experiment with smart grid technologies. Krotofil and
Larsen show in \cite{krotofil2015rocking} several testbeds and
simulations concluding that a successful attack against their
envisioned systems has to manage cyber and physical knowledge.

From a more control-theoretic standpoint, Candell et al. report
in \cite{candell2014cybersecurity} a testbed to analyze the
performance of security mechanisms for cyber-physical systems. The
work reports as well discussions from control and security
practitioners. McLaughlin et al. analyze
in \cite{mclaughlin2016cybersecurity} different testbeds and conclude
that it is necessary to use pathways between cyber and physical
components of the system in order to detect attacks. Also, Koutsandria
et al. \cite{koutsandria2015real} implement a testbed where the data
are cross-checked, using cyber and physical elements. Holm et al.
survey, classify and analyze in \cite{holm2015survey} several
cyber-physical testbeds proposed for scientific research. Inline with
the aforementioned contributions, we have presented in this paper an
ongoing testbed that aims at evaluating research mitigation
techniques targeting attacks at the physical layer of cyber-physical
systems operated via SCADA protocols. The initial focus of our testbed
has been the evaluation of the control-theoretic security mechanisms
reported in \cite{Mo_2015,Revisiting_Wat_based_detector,controlsys}.

\section{Conclusion}
\label{sec:conclusion}

In pursuance of security testing in cyber-physical systems, this paper
has provided a practical description of an ongoing platform to test
theoretical cyber-physical defense techniques. The architecture of the
testbed is based on real-world components, in order to emulate
cyber-physical systems commanded by SCADA (Supervisory Control And
Data Acquisition) technologies. Two real-world protocol implementations are included within our platform.

Three types of adversarial scenarios were also integrated in our testbed. The three scenarios enforce different types of attackers, incrementing the usability of the testbed to experiment novel security methods against a wider variety
of malicious intents. All three scenarios were confronted against
representative defense techniques. The platform also implements
testing automation in order to provide larger datasets as results and
enabling the architecture to perform repetitive tests. Experimental
results confirm previous theoretical and simulation-based work.\\

\noindent {\footnotesize {\bf Acknowledgements.} The authors
  acknowledge support from the Cyber CNI Chair of Institut
  Mines-T\'el\'ecom. The chair is held by T\'el\'ecom Bretagne and
  supported by Airbus Defence and Space, Amossys, EDF, Orange, La
  Poste, Nokia, Soci\'et\'e G\'en\'erale and the Regional Council of
  Brittany. It has been acknowledged by the Center of excellence in
  Cybersecurity.}

\end{document}